\documentclass[a4paper]{jpconf}
\usepackage{graphicx}
\usepackage{slashed}
\begin{document}
\title{Non-abelian T-duality, generalised geometry and holography.}

\author{N. T. Macpherson}

\address{Department of Physics, Swansea University, Singleton Park, Swansea, SA2 8PP, UK}

\ead{pymacpherson@swansea.ac.uk}

\begin{abstract}
Recent progress which relates non-abelian T-duality of $\mathcal{N}=1$ SuGra solutions to the powerful techniques of generalised geometry is reviewed. It is shown that $SU(3)$-structure solutions are mapped to $SU(2)$-structures and the transformation rule of the corresponding pure spinors is presented. This constitutes an important step towards utility of the duality within holography, showing for example, how smeared sources transform and so how flavour is included.
\end{abstract}
\section{Introduction}
Duality transformations play an important role in a wide range of physical systems from the 2d Ising Model to many dualities of superstring theory. Within the context of the gauge-gravity correspondence, duality transformation have in recent years proven to be a very useful way of generating new SuGra solutions dual to QFTs with phenomenologically desirable dynamics. An important example \cite{Maldacena:2009mw} is how U-duality can be used to map a deformation of the Maldacena-Nunez solution \cite{Maldacena:2000yy}, which gives a holographic description of a QFT with an irrelevant operator insertion, to the Baryonic branch of Klebanov-Strassler \cite{Butti:2004pk} which is well behaved asymptotically.

Here non-abelian T-duality is considered as a SuGra solution generator. Performing T-dualities along non-abelian isometries is not a new idea, indeed it was first proposed over 20 years ago \cite{de la Ossa:1992vc}. However, in order for it to have any real utility as a solution generating technique with applications to holography one first needed to know how non trivial RR fluxes transform under the duality. This was realised far more recently in \cite{Sfetsos:2010uq} and expanded upon in \cite{Lozano:2011kb}. 

Recently attention has been focused on generating new SuGra duals of $\mathcal{N}=1$ gauge theories in 4-d \cite{Itsios:2012zv} by dualising along the $SU(2)$ isometries of well established conifold solutions such as Klebanov-Witten \cite{Klebanov:1998hh}. Such solutions have a well understood generalised geometric description in terms of pure spinors \cite{Grana:2005sn} and it is has been shown, for a quite general ansatz, that the $SU(2)$ duality always maps between SuGra solutions \cite{Jeong:2013jfc}. This note reviews the results of \cite{Barranco:2013fza}, in which the techniques of non-abelian T-duality and generalised geometry are combined, constituting an important step towards constructing useful holographic duals of $\mathcal{N}=1$ QFT's in 4-d via the duality.
\section{Non-abelian T-duality along $SU(2)$ isometries}
The NS sector of a SuGra solution with an $SU(2)$ isometry can be generically expressed in terms of $SU(2)$ left invariant 1-forms, $L^i=-i Tr(g^{-1}dg)$ , as:
\begin{equation}
ds^2= G_{\mu\nu}dX^{\mu}dX^{\nu}+ 2G_{\mu i} dX^{\mu} L^i+ g_{ij} L^i L^j,~~~
B= B_{\mu\nu}dX^{\mu}\wedge dX^{\nu}+ B_{\mu i} dX^{\mu}\wedge L^i+ \frac{1}{2}b_{ij} L^i\wedge L^j
\end{equation}
where $\mu =1,...7$. The duality acts on the non linear sigma model corresponding to this solution
{\small\begin{equation}
S=\int d^2\sigma Q_{\mu\nu} \partial_+ X^{\mu}\partial_- X^{\nu}+Q_{\mu i} \partial_+X^{\mu}L_-^i+ Q_{i\mu}L_+^i \partial_- X^{\mu} +E_{ij}L^i_+ L^j_-,
\end{equation}}
where $Q=G+B$ and similarly for $E$. The dual sigma model is obtained by gauging the $SU(2)$ isometry by promoting $\partial_{\pm}$ to a covariant derivative in the pull back of the left invariant 1-forms, that is
{\small\begin{equation}
\partial_{\pm}g\to D_{\pm}g=\partial_{\pm}-A_{\pm} g.
\end{equation}}
A Lagrange multiplier $-i Tr(v F_{+-})$ should also be added to ensure the gauge field is non dynamical.

Upon integrating the Lagrange multiplier by parts the gauge fields may be solved for and and the T-dual sigma model obtained. Gauge fixing must be imposed as the dual theory depends on Euler angles of $SU(2)$ in addition to 3 dual coordinates $v_i$. The most simple fixing is $g=I$ in which only the $v_i$ are kept and one obtains the dual Lagrangian
{\small\begin{equation}
\tilde{S}=\int d^2\sigma Q_{\mu\nu} \partial_+ X^{\mu}\partial_+X^{\nu}+(\partial_+v_i+\partial_+ X^{\mu}Q_{\mu i})(E_{ij}+f_{ij}^{~~k}v_k)(\partial_-v_j -Q_{j \mu} \partial_- X^{\mu}),
\end{equation}}
from which the dual metric and NS 2-form may be extracted. In addition the dilaton obtains a correction at the quantum level
{\small\begin{equation}
e^{-2\hat{\Phi}}=det (E+fv)e^{-2\Phi}.
\end{equation}}

The transformation on the RR sector may be ascertained in a different fashion. After dualisation, left and right movers naturally couple to 2 distinct sets of vielbeins. However because these vielbeins describe the same geometry they must be related by a Lorentz transformation $\Lambda$. This may be used to define an action on spinors parametrised by a matrix $\Omega$ satisfying $\Omega^{-1}\Gamma^a\Omega =\Lambda^a_{~b}\Gamma^b$. For $SU(2)$ isometries this is solved by
{\small\begin{equation}
\Omega=\Gamma^{(11)}\frac{\Gamma^{123}+\sum_{a=1}^3\zeta^a \Gamma_a}{\sqrt{1 +\zeta^2}}
\end{equation}}
where $\zeta^a$ depends on the original geometry and the gauge fixing condition and can be found in \cite{Itsios:2012zv}. Dualising along odd dimensional isometrics maps type-IIA $\leftrightarrow$ type-IIB and the RR sectors of the original and dual theories are the related by
{\small\begin{equation}
e^{\Phi_{IIA}}\slashed{F}_{IIA} = e^{\Phi_{IIB}}\slashed{F}_{IIB}.\Omega^{-1}
\end{equation}}
where $F_{IIA/IIB}$ are RR polyforms and the slash denotes contraction with 10-d gamma matrices once the polyforms have been mapped to bispinors under the Clifford map.
\section{Generalised geometry and non-abelian T-duality}
As in \cite{Barranco:2013fza} the focus of this section shall be geometries that can be expressed as
\begin{equation}
ds^2_{str} = e^{2A}dx^2_{1,3} +ds^2_6,
\end{equation}
which preserve $\mathcal{N}=1$ and have non-trivial RR fluxes turned on. The MW 10-d killing spinors of such a solution my be written in terms of a 4+6 split
\begin{equation}
\epsilon^{1}=\zeta_+\otimes \eta^1_++\zeta_-\otimes \eta^1_-,~~~\epsilon^{2}=\zeta_+\otimes \eta^2_{\mp}+\zeta_-\otimes\eta^2_{\pm},
\end{equation}
where $\pm$ denotes chirality, the upper signs in $\epsilon^2$ corresponding to type-IIA, the lower to IIB and a basis such that $(\eta^{1,2}_+)^*=\eta^{1,2}_-$ is chosen. From the internal spinors (The $\eta$'s) it is possible to define 2 $Cliff(6,6)$ spinors
\begin{equation}
\Psi_{\pm}= \eta^1_+\otimes (\eta^2_{\pm})^{\dag}
\end{equation}
which map to polyforms under the Clifford map. The condition for $\mathcal{N}=1$ SUSY to hold may be caste in terms of differential constraints on these pure spinors
\begin{equation}
(d-H\wedge )(e^{2A-\Phi}\Psi_{\mp}) = e^{2A-\Phi}dA\wedge\bar{\Psi}_{\mp}+i\frac{e^{3A}}{8}\star_6 F_{IIA/IIB},~~~(d-H\wedge )(e^{2A-\Phi}\Psi_{\pm})=0
\end{equation}
where once more the upper signs should be taken in type-IIA.

A 6-d internal space which preserves $\mathcal{N}=1$ SUSY supports either an $SU(3)$ or an $SU(2)$-structure \cite{Grana:2005sn}. If $\eta^1_+$ is parallel to $\eta^2_+$, then the structure is $SU(3)$ and the pure spinors may be expressed as
\begin{equation}
\Psi_+=\frac{e^{A}}{8} e^{i \chi} e^{-iJ},~~~ \Psi_- =\frac{e^{A}}{8} \Omega_{hol}.
\end{equation}
where $\chi$ denotes a possibly point dependent phase. There exists a basis such that the projections of the background are canonical and $J$ and $\Omega_{hol}$ may be expressed in terms of 6-d vielbeins as
\begin{equation}
J=e^{12}+e^{34}+e^{56},~~~\Omega_{hol}=(e^1+i e^2)\wedge(e^3+i e^4)\wedge(e^5+i e^6).
\end{equation}
When $\eta^1_+$ is not parallel to $\eta^2_+$ each spinor defines an independent $SU(3)$-structure and the largest common subgroup is $SU(2)$. A special case is when the spinors bisect perpendicularly which gives an orthogonal $SU(2)$ characterised by pure spinors
\begin{equation}
\Psi_+ = \frac{e^{A}}{8} e^{-i v\wedge w}\wedge \omega,~~~ \Psi_-=ie^{i\xi}\frac{e^{A}}{8}(v+i w) e^{-i j}
\end{equation}
were $\omega$ is a 2-form and $v,w$ are 1-forms and $\xi$ is another phase.

The action of non-abelian T-duality on the 10-d spinors must change the chirality of $\epsilon_2$. When the original geometry is type-IIB the transformation is
\begin{equation}
\hat{\eta}^1_+ = \eta^1_+,~~~\hat{\eta}^2_+ = \Omega \eta^2_-,~~~ \hat{\eta}^{1,2}_-=(\tilde{\eta}^{1,2}_+)^*.
\end{equation}
This makes it abundantly clear that non-abelian T-duality is rotating the spinors with respect to each other and so one must map $SU(3)\to SU(2)$-structure under the duality. And as most known solutions of this type are $SU(3)$ structures, the duality provides a way of generating many more $SU(2)$-structure solutions, as long as the dualisation is performed on isometries that preserve SUSY.

The pure spinors transform under a non-abelian T-duality in much the same way as RR fluxes,
\begin{equation}
\slashed{\Psi}_{IIA\pm}= \slashed{\Psi}_{IIB\mp}\Omega^{-1}.
\end{equation}
As well as giving an explicit check of whether SUSY is preserved this also gives an immediate definition of the calibration conditions for space-time filling Dn-branes in the dual geometry.
\begin{equation} 
\sqrt{g}d^{n+1}\xi\bigg|_{\Sigma_{Dn}}=\Psi_{cal}\bigg|_{\Sigma_{Dn}},~~~\Psi_{cal}= -8e^{3A-\Phi} Im(\Psi_{\mp})
\end{equation}
where the upper sign is in IIA \cite{Martucci:2005ht}. This enables one to find SUSY brane embedding in the T-dual geometry which is important for adding flavour among other things. In addition smeared sources which enter a SuGra solution as a violation of the Bianchi identity of the RR polyforms
\begin{equation}
(d-H\wedge)F=\Xi\wedge e^{B}
\end{equation}
transform in precisely the same way as the RR sector. Together these facts allow one to generate new flavoured $SU(2)$-structure solution from flavoured $SU(3)$-structure solutions directly. 

In \cite{Barranco:2013fza} a first example of the utility of the techniques described above can be found. There KW \cite{Klebanov:1998hh} is dualised along one of its $SU(2)$ isometries. The resulting solution supports an orthogonal $SU(2)$-structure. The dualisation is repeated on KW with massless flavours \cite{Benini:2006hh} which generates sources for D8, D6 and D4 branes that are all SUSY. This constitutes a solid proof of concept.
\section{Outlook}
Holographic applications of non-abelian T-duality clearly require a greater understanding of the dual QFT's being generated. The large number of fluxes generated means that the duality probably maps to quivers, but this needs to be made more precise. The largest obstacle at this time is the periodicity of the dual coordinates. There is currently no canonical way to fix these and so many global properties of the dual manifolds are unknown. However one might reasonably hope that the generalised geometric picture may yet shed some light on this conundrum. 

$SU(2)$-structures are not that common in the literature so non-abelian T-duality could give some useful information about their general construction. Rarest of all is the dynamic $SU(2)$-structure where the angle between the internal spinors varies through out the geometry, it would be interesting to see if they can be generated with the duality.

Finally to what extent can generalised geometry help in the understanding non-abelian T-duals of other types types of solutions? For instance in \cite{Macpherson:2012zy} $G_2$-structure solutions dual to confining $\mathcal{N}=1$ Chern-Simons theories are considered. Does the duality map $G_2\to SU(3)$-structure in 7-d? This question is currently being addressed.

\ack
I would like to thank A.~Barranco, J.~Gaillard, C.~Nunez and D.~C.~Thompson for their collaboration on \cite{Barranco:2013fza} which this note is mostly based on. I am supported by an STFC studentship.

\end{document}